\documentstyle[12pt,epsfig,axodraw]{article}
\begin{document}
\baselineskip 0.7cm
%
%
\newcommand{\gsim}{ \mathop{}_{\textstyle \sim}^{\textstyle >} }
\newcommand{\lsim}{ \mathop{}_{\textstyle \sim}^{\textstyle <} }
\def\ul#1{\underline{#1}}
\def\J#1#2{J_{#1}\!\left(#2\right)}
%
\def\rc{r_c}
\def\pikrc{\pi{}k\rc}
\def\z1{e^{\pikrc}}
\def\zc{z_c}
\def\planck5d{M_{{\rm 5d}}}
\def\gauge5d{g_{{\rm 5d}}}
\def\yukawa5d{y_{{\rm 5d}}}
\def\quartic5d{\lambda_{{\rm 5d}}}
%
\def\L{{\rm L}}
\def\R{{\rm R}}
\def\psiL#1{\psi_\L^{(#1)}}
\def\psiR#1{\psi_\R^{(#1)}}
\def\psileft#1#2{\psiL{#1}\!\left(#2\right)}
\def\psiright#1#2{\psiR{#1}\!\left(#2\right)}
%
\def\leftmode#1#2{\xi_{#1}\!\left(#2\right)}
\def\rightmode#1#2{\eta_{#1}\!\left(#2\right)}
\def\gaugemode#1#2{{\widehat \chi}_{#1}\!\left(#2\right)}
\def\zeromode#1{\widehat{\zeta}\!\left(#1\right)}
\def\whleftmode#1#2{{\widehat \xi}_{#1}\!\left(#2\right)}
\def\whrightmode#1#2{{\widehat \eta}_{#1}\!\left(#2\right)}
%
\def\gaugemass#1{M_{{#1}}}
\def\eigen#1{\lambda_{#1}}
\begin{titlepage}
\begin{flushright}
KEK-TH-690\\
March 2000
\end{flushright}

\begin{center}
{\large \bf Proton Decay in the Supersymmetric Grand Unified Models}

\vskip 1.2cm

Junji~{\sc Hisano}%

\vspace{10mm}

 {\it Theory Group, KEK, Tsukuba, Ibaraki 305-0801, Japan\\
        Email:junji.hisano@kek.jp}
\vskip 1.5cm

\abstract {
In this article we review proton decay in the supersymmetric grand
unified models.
}
\end{center}

\setcounter{footnote}{0}
\end{titlepage}

\section{Introduction}

The grand unified model (GUT) of the strong and electroweak
interactions is motivated not only from an esthetic point of view, but
also due to the problem of the electric charge quantization in the
standard model (SM) \cite{PRL32-438}.  The electric charge
quantization is experimentally proved up to $10^{-21}$
\cite{EPJC3-794}.  Also, the unification of the gauge coupling
constants, which is one of the GUT predictions, is shown to be
valid in the supersymmetric extension of the GUT (SUSY GUT), and it is
found that the GUT scale ($M_{\rm GUT}$) is about $10^{16}$ GeV
\cite{PRD44-817}.

The GUT has two aspects. One is the unification of the SM gauge groups
to a simple group, such as SU(5) or SO(10).  The charge quantization
and the gauge coupling constant unification come from the unification
of gauge groups. The other is unification of matters. In the SU(5)
GUT, quarks and leptons are embedded in an economical and elegant way
as
\begin{eqnarray}
\psi({\bf 10}) = 
\left(
\begin{array}{ccc}
Q,& u^{\cal C},& e^{\cal C}
\end{array}
\right), \nonumber\\
\phi({\bf 5}^\star) = 
\left(
\begin{array}{cc}
d^{\cal C}, &L
\end{array}
\right), \nonumber
\end{eqnarray}
where $Q\equiv(u,d)$ and $L\equiv(\nu,e)$.  Here, we adopt the
left-handed basis for fermions and omit the chirality indices of
fermions as far as it does not lead to confusion in this article. In
the SO(10) GUT the right-handed neutrinos are introduced and the
embedding becomes simpler as
\begin{eqnarray}
\varphi({\bf 16}) = 
\left(
\begin{array}{cccc}
u^{\cal C},& d^{\cal C},& e^{\cal C},& \nu^{\cal C} \\
u  ,& d  ,& e  ,& \nu 
\end{array}
\right). \nonumber
\end{eqnarray}
The atmospheric neutrino observation by the Super-Kamiokande experiment
\cite{PRL81-1562} suggests that the tau neutrino has a finite mass
($m_{\nu_\tau} \sim 10^{-(1-2)}$eV). If the tiny neutrino mass comes
from the see-saw mechanism \cite{YGRS}, the right-handed tau neutrino
mass is expected to be ($10^2$GeV)$^2$/$10^{-(1-2)}$eV$\sim
10^{15}$GeV. It is close to the GUT scale and supports the SO(10)
SUSY GUT.

The unification of matters leads to violation of the global
symmetries in the SM, such as baryon, lepton, and lepton flavor
numbers, and then, proton decay and lepton-flavor violating processes
are predicted.  In this article we will concentrate on the proton decay in
the SUSY SU(5) GUT. The proton decay is the direct prediction for the
model, and the search is the most important for confirmation of the
GUT. For the lepton-flavor violating processes in the SUSY
GUT, see Refs.~\cite{PLB338-212}.

There are two sources to induce the proton decay in the SUSY SU(5)
GUT. One is the $X$ boson, accompanied with the unification of gauge
groups.  The proton lifetime predicted by the $X$ boson exchange is
proportional to $M_X^4$ with $M_X$ the $X$ boson mass, since the $X$
boson exchange induces the dimension-six operators.  The dominant
decay mode is $p\rightarrow \pi^0 e^+$.  While this prediction is
almost model-independent, the lifetime is sensitive to $M_X$. If
$M_X\sim 10^{16}$GeV, the lifetime is $10^{(34-36)}$ years, which is
beyond the current experimental reach.
  
The second one is the colored Higgs, which is introduced for doublet
Higgs in the SUSY SM to be embedded into the SU(5) multiplets.  In the
minimal SUSY SU(5) GUT, the colored Higgs predicts much shorter proton
lifetime. The colored Higgs exchange leads to the baryon-number
violating dimension-five operators, and then the proton lifetime is
proportional to $M_{H_C}^2$ with $M_{H_C}$ the colored Higgs mass
\cite{NPB197-533,PRD32-2348,NPB402-46}.  While it is
suppressed by the small Yukawa coupling of quarks and leptons, the
minimal SUSY SU(5) GUT is excluded from the negative search of the 
proton decay \cite{PRD59-115009}.

This paper is organized as follows.  In the next section, we show the
status of the minimal SUSY SU(5) GUT from a point of view of the proton
decay induced by the colored Higgs, and discuss the reason why the
proton decay is not still discovered if the SUSY SU(5) GUT is valid. In
Section 3 we show the proton decay rate induced by the $X$ boson, and
discuss the model-dependence of the prediction. Section 4 is devoted to
conclusion. 

\section{Proton decay induced by the colored Higgs exchange}

In this section we show the status of the minimal SUSY SU(5) GUT from
a point of view of the proton decay induced by the colored Higgs, and
discuss the reason why the proton decay is not discovered if the SUSY 
SU(5) GUT is realistic.

First, let us introduce the minimal SUSY SU(5) GUT. In this model the
doublet Higgs $H_f$ and $\overline{H}_f$ in the SUSY SM are embedded
into the {\bf 5} and {\bf 5$^\star$} dimensional multiplets as
\begin{eqnarray}
H =(H_C,\,H_f), &&
\overline{H} =(\overline{H}_C,\,\overline{H}_f),
\end{eqnarray}
with the colored Higgs $H_C$ and $\overline{H}_C$. The superpotential
of the Yukawa coupling for the doublet and colored Higgs is given as 
\begin{eqnarray}
W_Y &=&  h^i \ (Q_i \cdot H_f)   u_i^{\cal C} 
      +  V_{ij}^\ast f^j \ (Q_i \cdot \overline{H}_f)   d_j^{\cal C}  
      + f^i \  e_i^{\cal C}  (L_i \cdot \overline{H}_f)
     \nonumber
      \\
   && +  \frac12 h^i e^{i \varphi_i} \ (Q_i \cdot  Q_i)   H_C
      +  V_{ij}^\ast f^j \ (Q_i \cdot  L_j)  \overline{H}_C
     \nonumber\\
   && +  h^i  V_{ij} \ u_i^{\cal C} e_j^{\cal C}   H_C
      +  e^{ - i \varphi_i} V_{ij}^\ast f^j \  u_i^{\cal C} d_j^{\cal C}  \overline{H}_C,
\label{yukawa}
\end{eqnarray}
where the indices $i$ and $j(=1-3)$ are for the generations. $V_{ij}$
 is the Kobayashi-Maskawa matrix element at the GUT scale, and
 $\varphi_i$ are for additional degrees of freedom in the Yukawa
 coupling constants in the minimal SUSY SU(5) GUT. (See
 Ref.~\cite{NPB402-46} for notation and conversion.) Here, we use the
 same letters for the superfields as the component fields. The
 first-three terms correspond to the Yukawa coupling in the SUSY
 SM. As well-known, the charged leptons and the down-type quarks have
 common Yukawa coupling constants.

In the minimal SUSY SU(5) GUT $H_C$ and $\overline{H}_C$ have a common
 mass term by themselves, and then, the colored Higgs exchange gives
 following dimension-five operators,
\begin{equation}
W_5 = \frac{1}{2M_{H_C}} h^{i} e^{i \varphi_i} V_{kl}^\ast f^l
	\ (Q_i \cdot Q_i) (Q_k \cdot L_l)
    + \frac{1}{M_{H_C}} h^i V_{ij} e^{-i \varphi_k} V_{kl}^\ast f^l
        \ u^{\cal C}_i e^{\cal C}_j u^{\cal C}_k d^{\cal C}_l.
		\label{dimen5}
\end{equation}
Since squarks or sleptons are on the external lines of these
operators, Higgsino or gaugino is exchanged between them so that
proton can decay (Fig.~(1)).  
\begin{figure}[t]
\begin{center} 
\begin{picture}(200,190)(0,10) 
\Line(30,30)(70,100) \Text(20,30)[t]{$q$}
\Line(30,170)(70,100)  \Text(20,170)[t]{$l$}
\DashLine(145,70)(70,100){5}  \Text(105,130)[t]{$\tilde{q}$}
\DashLine(145,130)(70,100){5} \Text(105,70)[t]{$\tilde{q}$}
\Line(145,70)(145,130)   \Text(160,100)[t]{$\tilde{g},~\tilde{h}$}
\Line(145,70)(170,30)     \Text(180,30)[t]{$q$}
\Line(145,130)(170,170)    \Text(180,170)[t]{$q$}
\end{picture}
\end{center}
\caption{Higgsino and gaugino dressing to the baryon-number violating dimension-five operators.}
\end{figure}
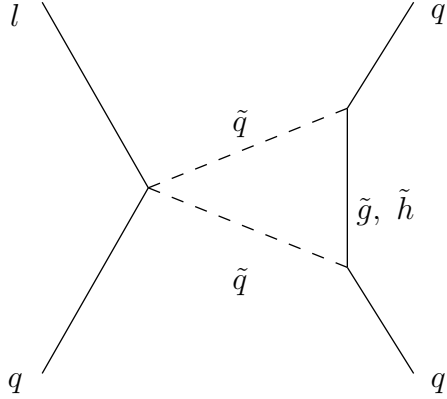
The contractions of the indices for the gauge symmetries in
Eq.~(\ref{dimen5}) are understood as
\begin{eqnarray}
	(Q_i \cdot  Q_i) (Q_k\dot  L_l) &=& \varepsilon_{\alpha \beta \gamma}
		(u_i^\alpha d_i^{\beta} -
			d_i^{\alpha} u_i^\beta )
		(u_k^\gamma e_l - d_k^{\gamma} \nu_l), \\
	u^{\cal C}_i e^{\cal C}_j u^{\cal C}_k d^{\cal C}_l &=& \varepsilon^{\alpha \beta \gamma}
		u^{\cal C}_{i\alpha} e^{\cal C}_j u^{\cal C}_{k\beta} d^{\cal C}_{l\gamma},
\end{eqnarray}
with $\alpha$, $\beta$, and $\gamma$ being color indices. Note that
the total antisymmetry in the color indices requires that the
operators are flavor non-diagonal $(i \neq k)$. Therefore the dominant
decay modes tend to have strangeness. In the minimal SUSY SU(5) GUT
the dominant mode is $p\rightarrow K^+ \bar{\nu}$ while $p\rightarrow
K^0 \mu^+$ is not. The mode of $K^0 \mu^+$ is suppressed by $(m_u/m_c
V_{ud}V_{cd})^2$.

The largest uncertainty to the decay rate predicted by the colored
Higgs exchange comes from the colored Higgs mass itself in the minimal SUSY
SU(5) GUT. However, the mass spectrum at the GUT scale can be
constrained and the colored Higgs mass can be evaluated from the low
energy parameters \cite{PRL69-1014}. Due to the gauge coupling
unification in the minimal SUSY SU(5) GUT we can derive the following
relations from the renormalization group equations of the gauge
coupling constants at one-loop level,
\begin{eqnarray}
(- 2 \alpha_3^{-1} + 3 \alpha_2^{-1} - \frac35 \alpha_Y^{-1}) (m_Z)
	&=& \frac{1}{2\pi} \left\{
		\frac{12}{5} \, \ln \frac{M_{H_C}}{m_Z}
		- 2 \, \ln \frac{m_{SUSY}}{m_Z} \right\},
			\label{MHC}
			\\
(- 2 \alpha_3^{-1} - 3 \alpha_2^{-1} + 3 \alpha_Y^{-1} ) (m_Z)
	&=& \frac{1}{2\pi} \left\{
		12 \, \ln \frac{M_X^2 M_\Sigma}{m_Z^3}
		+ 8 \ln \frac{m_{SUSY}}{m_Z} \right\}.
		\label{MGUT}
\end{eqnarray}
Here, we take the SUSY particle masses in the SUSY SM to be common
($m_{SUSY}$).  $M_\Sigma$ is the mass for the SU(3)$_C$ octet and
SU(2)$_L$ triplet components in the {\bf 24}-dimensional Higgs for
SU(5) gauge symmetry breaking, which survive as the physical degrees
of freedom after the Higgs mechanism. In Ref.~\cite{MPLA10-2267} the
upperbound on $M_{H_C}$ is derived from Eq.~(\ref{MHC}), including the
various correction to it, as
\begin{equation}
M_{H_C} \le 2.4 \times 10^{16}{\rm GeV}.
\end{equation}
On the other hand, the recent study for the proton decay
\cite{PRD59-115009} shows
\begin{equation}
M_{H_C} \ge 7.2 \times 10^{16}{\rm GeV}
\end{equation}
from the experimental bound on the proton lifetime $\tau(p\rightarrow
K^+ \bar{\nu}) > 6.7\times 10^{32}$ years \cite{PRL83-1529}, assuming
the squark and slepton masses are lighter than 1TeV. Then, the minimal
SUSY SU(5) GUT is excluded now.
 
While the minimal SUSY SU(5) GUT is excluded, the prediction of the
proton decay induced by the colored Higgs exchange is itself highly
model-dependent. Let us discuss the model-dependence. First, though it
depends on the Yukawa coupling structure at the GUT scale, we cannot
explain the mass relations between the down-type quarks and the charge
leptons in the first and second generations in an SU(5) symmetric way
as Eq.~(\ref{yukawa}).  We need to add some modification to it. One of
the examples is introduction of the higher dimensional operators
suppressed by the gravitational scale.  If the Yukawa coupling of the
colored Higgs is suppressed automatically or accidentally, the
lowerbound on $M_{H_C}$ becomes looser. In this case, the mode of $K^0
\mu^+$ may be dominant. One of the explicit models is given in
Ref.~\cite{PAN60-126}.

Second, in some models the effective mass for the colored Higgs, which
suppressed the dimension-five operators, can be much heavier than the
GUT scale \cite{PLB291-263,PLB342-138,PRD48-5354}. The dimension-five operators
given in Eq.~(\ref{dimen5}) are induced since $H_C$ and
$\overline{H}_C$ have a common mass term. If $H_C$ and
$\overline{H}_C$ have mass terms independent of each others by
introduction of $H_C^\prime$ and $\overline{H}_C^\prime$, the
dimension-five operators are not induced. 

If the Peccei-Quinn (PQ) symmetry \cite{PRL38-1440} is introduced,
this mechanism works \cite{PLB291-263,PLB342-138}. The PQ charges for matters and
Higgs are assigned as
\begin{eqnarray}
Q_{\rm PQ}(\psi_i) = 1, &&Q_{\rm PQ}(\phi_i) = 1, \nonumber\\
Q_{\rm PQ}(H) = -2, && Q_{\rm PQ}(\overline{H}) = -2, \nonumber\\
Q_{\rm PQ}(H^\prime) = 2, && Q_{\rm PQ}(\overline{H}^\prime) = 2, \nonumber
\end{eqnarray}
where $H^\prime(\equiv(H_C^\prime,H_f^\prime))$ and
$\overline{H}^\prime(\equiv(\overline{H}_C^\prime,\overline{H}_f^\prime))$.
This symmetry prohibits the mass term $\sim M_{H_C} \overline{H}_C
H_C$ and the dimension-five operators in Eq.~(\ref{dimen5}). After the
PQ symmetry is broken at $M_{\rm PQ}\sim 10^{(10-12)}$ GeV, the
dimension-five operators are generated.  However, they are suppressed
by $M_{\rm PQ}/M_{\rm GUT}\sim 10^{-(4-6)}$ and they are harmless.
 
In order to confirm the SUSY GUT, we need more model-independent
predictions, which come from the gauge sector.  One of them is the
proton decay induced by the $X$ boson. Then, we would like to discuss the
model-dependence in next section. 

\section{Proton decay induced by the $X$ boson exchange}

In this section we show the proton decay rate induced by the $X$ boson
exchange, and discuss the model-dependence.  The $X$ boson,
is SU(3)$_C$ ${\bf 3^\star}$- and SU(2)$_L$ {\bf 2}-
dimensional, and the hypercharge is ${\bf -5/6}$. The
interaction of the $X$ boson to matter fermions is given as follows;
\begin{eqnarray}
{\cal L} &=& -\frac{1}{\sqrt{2}} g_5 V_{ij}   \
	\overline{d_{j}^{\cal C}} (X_\mu \cdot \gamma^\mu L_{i})
 +	\frac{1}{\sqrt{2}} g_5 {\rm e}^{-i \varphi_i} \
	\overline{Q_{i}} X_\mu \gamma^\mu u_{i}^{\cal C}
\nonumber\\
&&
 +	\frac{1}{\sqrt{2}} g_5 V_{ij} \
	(\overline{e_{j}^{\cal C}} \cdot X_\mu) \gamma^\mu Q_{i}
 + h.c..
\end{eqnarray}
This interaction leads to the following baryon-number violating
 operators, which contribute to $p\rightarrow \pi^0 e^+$,
\begin{eqnarray}
{\cal L}_{eff} &=& 
		    A_R \frac{g_5^2}{M_X^2} {\rm e}^{i \varphi_1} \times
\nonumber\\
&&
		   \epsilon_{\alpha\beta\gamma} 
	           (
	           (\overline{d_L^{\cal C}})^\alpha (u_R)^\beta
	           (\overline{u_R^{\cal C}})^{\gamma} (e_L)
	          +(1+|V_{ud}|^2)
	           (\overline{d_R^{\cal C}})^\alpha (u_L)^\beta
	           (\overline{u_L^{\cal C}})^{\gamma} (e_R)
		    ) 
\label{dim6}
\end{eqnarray}
where $A_R$ is the renormalization factor from the anomalous
dimensions to these operators. 

The renormalization factor $A_R$ has the short-distance contribution
($A_R^{(SD)}$) and the long-distance contribution
($A_R^{(LD)}$). $A_R^{(SD)}$ at one-loop level is given as
\begin{eqnarray}
A_R^{(SD)} &=&
\left(\frac{\alpha_3(m_Z)}{\alpha_5}\right)^{\frac{4}{3 b_3}} 
\left(\frac{\alpha_2(m_Z)}{\alpha_5}\right)^{\frac{3}{2 b_2}} 
\label{ar_sd}
\end{eqnarray}
where $b_3= 9-2n_g$ and $b_2= 5-2 n_g$ with $n_g$ the number of the
generations \cite{NPB245-425}. Here, we do not include the U(1)$_Y$
contribution. It is expected to be smaller while it has not been
calculated completely. If the SUSY SM is the low energy effective theory
below the GUT scale, $A_R^{(SD)}=2.1$ for $\alpha_3(m_Z) = 0.116$,
$\sin^2\theta_W = 0.2317$, and $\alpha^{-1}(m_Z) = 127.9$.
The long-distance part $A_R^{(LD)}$ is
\begin{eqnarray}
A_R^{(LD)} &=&
\left(\frac{\alpha_3(m_b)}{\alpha_3(m_Z)}\right)^{\frac{6}{23}} 
\left(\frac{\alpha_3(\mu_{had})}{\alpha_3(m_b)}\right)^{\frac{6}{25}}
= 1.2. 
\end{eqnarray}

From the effective lagrangian (\ref{dim6}) the proton lifetime from
$p\rightarrow \pi^0 e^+$ is
\begin{eqnarray}
\Gamma(p\rightarrow \pi^0 e^+) &=& \alpha_H^2 \frac{m_p}{64 \pi f_\pi^2}
(1+D+F)^2 \left(\frac{g_5^2}{M_X^2} A_R \right)^2 (1+ (1+|V_{ud}|^2)^2),
\end{eqnarray}
where $m_p$ is the proton mass, $f_\pi$ is the pion decay constant,
and $D$ and $F$ are the chiral lagrangian parameters. The definition
of $\alpha_H$ is
\begin{eqnarray}
\alpha_H u_L({\bf k}) 
&\equiv& 
\epsilon_{\alpha\beta\gamma} 
\langle 0|
(\overline{d_L^{\cal C}})^\alpha (u_R)^\beta
(\overline{u_R^{\cal C}})^{\gamma} 
|p_{\bf k} \rangle.
\end{eqnarray}
$\alpha_H$ had one-order ambiguity in the old theoretical calculations
and it led to a large uncertainty to the prediction.  The latest
evaluation by lattice calculation \cite{hep-lat/9911026} gives 
\begin{eqnarray}
\alpha_H&=&-(0.015\pm 1) {\rm GeV^3}.
\end{eqnarray}
They adopt the naive dimensional renormalization scheme with the
renormalization scale ($\mu_{had}$) 2.3GeV. This lattice calculation
is the first realistic one, in which enough a large lattice spacing
and a large statistics are prepared.  While this calculation still has
ambiguities from the quenched approximation and the $a^{-1}$
correction, they are expected to be at most $O(10)$\%. Finally, we get
\begin{eqnarray}
1/\Gamma(p\rightarrow \pi^0 e^+) 
&=&1.0 \times 10^{35} {\rm years} 
\nonumber\\
&&
\times \left(\frac{\alpha_H}{0.015 {\rm GeV}^3}\right)^{-2}
\left(\frac{\alpha_5}{1/25}\right)^{-2}
\left(\frac{A_R}{2.5}\right)^{-2} 
\left(\frac{M_X}{10^{16} {\rm GeV}}\right)^4 .
\end{eqnarray}

\begin{figure}[t]
\begin{center}
\centerline{\leavevmode\psfig{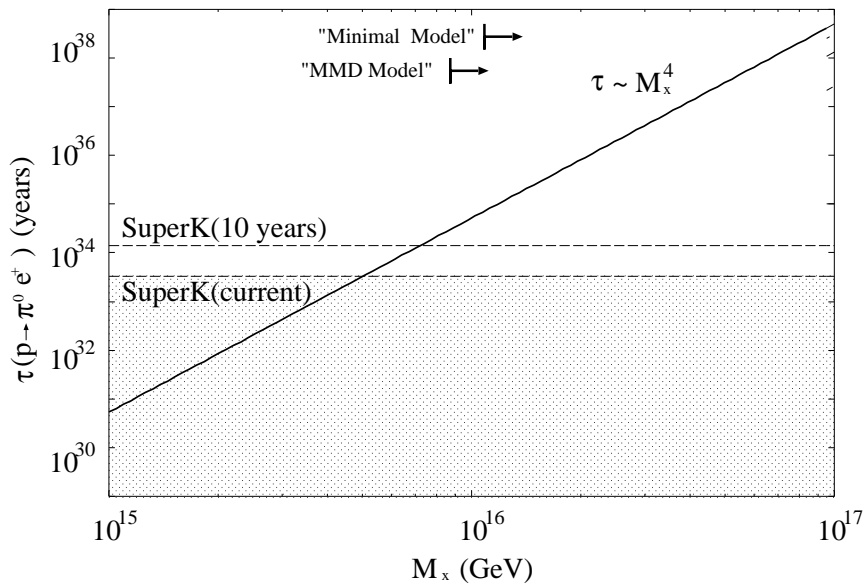}}
\end{center}
\caption{Lifetime of proton as a function of $M_X$. The shaded region
has been excluded by Super-Kamiokande experiment ($ 3.3 \times
10^{33}$ years).  The dash-doted line is the expected reach of ten
years run of the Super-Kamiokande experiment ($ 1.4 \times 10^{34}$
years) \cite{shiozawa-nnn}.  }
\label{fig.2}
\end{figure}

In Fig.~(2) we show the lifetime of proton as a function of $M_X$.
The shaded region has been excluded by Super-Kamiokande experiment.
The dash-doted line is the reach of ten years run of the
Super-Kamiokande experiment. In Fig.~(2) we show the lowerbound on the
$X$ boson mass in the minimal SUSY SU(5) GUT as a reference point
\cite{NPB402-46},
\begin{equation}
M_X \ge 1.1 \times 10^{16}{\rm GeV}.
\end{equation}
This bound comes from the constraint from the gauge
coupling unification given in Eq.~(\ref{MGUT}),
\begin{equation}
1.3\times10^{16}{\rm GeV} \le ({M_X^2 M_\Sigma})^{\frac13}
\le 3.2\times10^{16}{\rm GeV}
\end{equation}
and validity of perturbation below the gravitational scale,
\begin{equation}
M_{\Sigma}/M_{X} = \lambda_{\Sigma}/2 \sqrt{2} g_5 \le 1.8.
\end{equation}
$\lambda_\Sigma$ the self-coupling constant of the {\bf
24}-dimensional Higgs multiplet, and if $\lambda_\Sigma$ is much
larger than $g_5$, it blows up below the gravitational
scale. Similarly, we can derive the lowerbound on $M_X$ in the
Modified Missing Doublet (MMD) model as $M_X \ge 8.7 \times
10^{15}$GeV \cite{PTP98-1385}. From this figure, it is found that we
need further effort to confirm or reject the SUSY GUT.

We have assumed that the SUSY SM is valid below the GUT
scale. However, even if extra SU(5) complete multiplets have smaller
masses than the GUT scale, the success of the gauge coupling unification
is not spoiled. In fact, some models, such as the gauge mediated SUSY
breaking model \cite{PRD53-2658}, the anomalous U(1) SUSY breaking
model given in Ref.~\cite{PLB445-316}, and the E$_6$ model
\cite{PREP183-193}, predict existence of the extra matters at lower
energy. In this case the the gauge coupling constant at the GUT scale
becomes larger. Also, the renormalization factor $A_R$ is
enhanced as Eq.~(\ref{ar_sd}). Then, for fixed $M_X$, the proton lifetime 
becomes shorter.

\begin{figure}[t]
\begin{center}
\centerline{\leavevmode\psfig{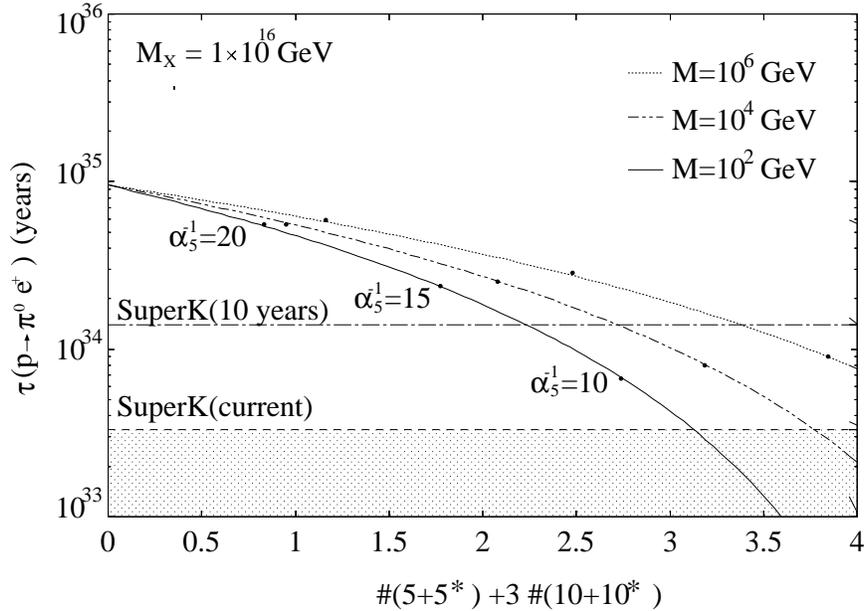}}
\end{center}
\caption{Lifetime of proton as a function of contribution to the beta
functions of the gauge coupling constants from the extra
matters. Here, we take $M_X=1.0\times 10^{16}$ GeV.  The masses for the
extra matters $M$ are taken to be $10^2$, $10^4$, and $10^6$ GeV.}
\label{fig.3}
\end{figure}

In Fig.~(3) we show the proton lifetime as a function of
contribution to the beta functions of the gauge coupling constants
from the extra matters. The contribution from a pair of {\bf 5} and
{\bf 5$^\star$} is one, and that from a pair of {\bf 10} and {\bf
10$^\star$} is three. Here, we fix $M_X=1.0\times 10^{16}$ GeV.  and
take the extra matter mass $M$ to be $10^2$, $10^4$, and $10^6$ GeV.
From this figure, introduction of two extra generations (a pair of
{\bf 5} and {\bf 5$^\star$} and a pair of {\bf 10} and {\bf 10$^\star$})
with the masses $10^2$ GeV is slightly disfavored.

\section{Conclusion}

The proton decay search is the most important for confirmation of the
GUT.  Now the minimal SUSY SU(5) GUT is excluded by the negative
search since the dimension-five operators induced by the charged Higgs
exchange predict a large proton decay rate. However, the proton decay
by the colored Higgs is highly model-dependent, and it is premature to
conclude that the SUSY GUT is excluded. We hope that the search will be
pushed as far as possible.

\section*{Acknowledgement}

We appreciate T. Onogi and N. Tsutsui for useful comments.

\newcommand{\Journal}[4]{{#1} {#2} {(#3)} {#4}} \newcommand{\PL}{Phys. 
Lett.}  \newcommand{\PR}{Phys. Rev.}
\newcommand{\PRL}{Phys. Rev. Lett.}  \newcommand{\NP}{Nucl. Phys.}
\newcommand{\ZP}{Z. Phys.}  \newcommand{\PTP}{Prog. Theor. Phys.}
\newcommand{\NC}{Nuovo Cimento} \newcommand{\MPL}{Mod. Phys. Lett.}
\newcommand{\PRep}{Phys. Rep.}

\end{document}